# Detecting Carbon Nanotube Orientation with Topological Data Analysis of Scanning Electron Micrographs

Liyu Dong[1, 2, †] Haibin Hang [3, *, †], Jin Gyu Park[1, 4,*], Washington Mio[3], Richard Liang[1, 4]

[1]High-Performance Materials Institute (HPMI), Florida State University, 2005 Levy Ave., Tallahassee, FL 32310, USA

[2]Materials Science and Engineering, Florida State University, 2005 Levy Ave., Tallahassee, FL 32310, USA

[3]Department of Mathematics, Florida State University, 208 Love Building, 1017 Academic Way, Tallahassee, FL 32306, USA

[4]Department of Industrial and Manufacturing Engineering, FAMU-FSU College of Engineering, 2525 Pottsdamer St., Tallahassee, FL 32310, USA

[†] These authors declare equal contributions.

* Corresponding authors: hhang@math.fsu.edu (H. Hang) jgpark@fsu.edu (J.G. Park)

## ABSTRACT

As the aerospace industry becomes increasingly demanding for stronger lightweight materials, the ultra-strong carbon nanotube (CNT) composites with highly aligned CNT network structures could be the answer. In this work, a novel methodology applying topological data analysis (TDA) to the scanning electron microscope (SEM) images was developed to detect CNT orientation. The CNT bundle extensions in certain directions were summarized algebraically and expressed as visible barcodes. The barcodes were then calculated and converted into the total spread function $V(X,\theta)$, from which the alignment fraction and the preferred direction could be determined. For validation purposes, the random CNT sheets were mechanically stretched at various strain ratios ranging from 0-40%, and quantitative TDA analysis was conducted based on the SEM images taken at random positions. The results showed high consistency ($R^2$=0.975) compared to the Herman's orientation factors derived from the polarized Raman spectroscopy and wide-angle X-ray scattering analysis. Additionally, the TDA method presented great robustness with varying SEM acceleration voltages and magnifications, which might alter the scope in alignment detection. With potential applications in nanofiber systems, this study offers a rapid and simple way to quantify CNT alignment, which plays a crucial role in transferring the CNT properties into engineering products.

## 1. INTRODUCTION

The efficient transfer of the exceptional mechanical [1] and electrical [2] properties of carbon nanotube (CNT) materials into nanocomposite systems requires excellent quality [3], adequate alignment, uniform dispersion and strong interfacial bonding [4]. Efforts to overcome the inter-nanotube van der Waals forces and induce bundle orientation include filtration and drawing [5], magnetic field alignment [6], shear pressing or mechanical densification [7, 8] and mechanical stretching [9, 10], revealing a strong correlation between alignment fractions and resultant mechanical and electrical properties. Han *et al*. [4] reported the tensile strengths of the CNT/bismaleimide composites over 6 GPa using multi-step stretching, shedding light on furthering the rapid development of CNT composites as a promising replacement for the prevalent carbon fiber reinforced plastics (CFRPs).

Despite various efforts implemented [11-13], the nanoscale anisotropy and alignment mechanisms have yet to be fully modelled or understood, necessitating the development of reliable representations to precisely quantify the bundle orientation. X-ray scattering and polarized Raman spectroscopy are currently widely accepted to analyze the nanotube bundle orientation: X-ray scattering provides the Fourier transform of in-situ atomic space information, which can be applied to investigate the CNT packing and orientation [6, 14]; using a polarized laser source, one can obtain the distribution functions that are characteristic of CNT orientation by fitting the Raman spectra at various angles between the CNT axis and the incident excitation polarization [15]. Recently, Brandley *et al*. reported a novel method of mapping the CNT alignment with the optimized SEM imaging parameters using two-dimensional Fourier transform, and demonstrated a narrower distribution function and higher Herman parameter as pre-strain increases [16].

In this study, we explored the use of topological data analysis (TDA) to effectively detect CNT bundle orientation based on SEM images. Years of rapid development has led to the blooming of TDA, which offers solutions to complex image and data analysis problems in diverse interdisciplinary applications [17-19]. This approach requires extensive analysis of the topological features of data [20, 21], such as connected components (0-dim), loops (1-dim), and cavities (2-dim). As a fundamental TDA tool, the persistent homology (PH) measures how the topological features evolve in a 1D-parameterized data organization. Visible barcode expressions of PH summarize the birth and death time of each feature along the 1D-parameter. Significant theoretical work has shown that the barcode expressions are robust even under small perturbations in the input data [22], and efficient algorithms were accordingly formulated to compute the barcode summation of topological features [23].

We have quantified the CNT bundle orientation using persistent homology, where the data features (i.e., the CNT bundles extensions in certain directions) were summarized in an algebraic way and expressed in visible barcodes. An easy observation is that the nanotube bundles at different depths tend to overlap or intersect with each other, and they spread or expanded longer in the alignment direction, as depicted in Figure 1. Particularly, we scanned the SEM images of CNT bundle network back and forth in any given direction, which generated two different 1D-parameterized organization of each image, known as the sub-level set filtration and super-level set filtration of the projection map. Innovatively, we have combined the information encoded in the barcodes of both filtrations to demonstrate the total CNT bundles variations more precisely. One of the most prominent advantages of this work is that the total extensions of all bundles in an SEM image were summarized and analyzed all together to demonstrate the overall alignment fraction. An example Python code underlying our TDA approach is publicly available at https://github.com/Haibin9632/cnt_tda.

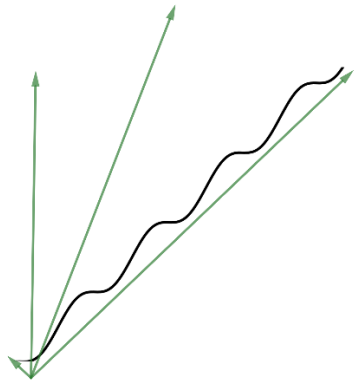

Figure 1. A schematic representation of an individual CNT bundle (black curve) with each vector (in green) showing the bundle extensions in the corresponding direction.

## 2. Experimental

The randomly oriented CNT sheets supplied by Nanocomp Inc (Concord, NH, USA), were mechanically stretched to varying ratios using a continuous stretching approach as previously reported [24, 25], which were denoted as 0%, 10%, 20%, 30%, and 40%, based on the width changes. With the average CNT diameters being 6-10 nm, the primarily double-walled CNTs were comprised of bundles with an aspect ratio of ~100,000. Approximately 10-15 wt% residue iron was revealed by thermogravimetric analysis [25].

Assuming a uniform alignment distribution, both random and aligned CNT tapes were examined using scanning electron microscopy (SEM, JSM-7401F, JEOL) for morphology analysis at different magnifications (×5,000, ×10,000, ×15,000, and ×20,000) under different accelerating voltages (5 kV and 10kV), with the resolution set at 1024×1280 pixels. To evaluate the efficiency and accuracy of our algorithms, the polarized Raman spectroscopy and wide-angle X-ray scattering (WAXS) were performed. With an excitation wavelength of 785 nm (1.58 eV) and objective lens of 50× to regulate spot size within 1 μm, the Renishaw inVia confocal micro-Raman system with a typical 0.5 mW laser power was used. The polarized Raman spectra were collected in the VV configuration [9, 10], where the incident and scattering light were polarized along the preferred direction. For thin CNT sheets (<50 μm), a two-dimensional model was constructed by neglecting the anisotropic laser penetration depth [6], and the degree of alignment was calculated by quantifying the deviation from the perfect alignment scenario, following

$$I(\varphi, f, \sigma) = A \int_0^{\frac{\pi}{2}} [\frac{1-f}{\pi} + \frac{f}{\sigma\sqrt{\pi/2}} e^{-2(\theta-\varphi)^2/\sigma^2}] \cdot \frac{\cos^4 \theta}{\cos \theta + K \sin \theta} \, d\theta \qquad (1)$$

where $f$ is the alignment degree and $\sigma$ the Gaussian standard deviation, which is equivalent to the full width at half maximum (FWHM).

The wide-angle X-ray scattering (WAXS) measurements were performed on a Bruker NanoSTAR system, with an Incoatec IμS microfocus X-ray source operating at 45 kV, 650 μA. The primary beam was collimated with cross-coupled Göbel mirrors and a 3-pinhole collimation system, providing a Cu Kα radiation beam (λ = 0.154 nm) with a beam size about 0.15 mm at the sample position. The wide-angle pattern and intensity were captured by a photo image plate and read with a FLA-7000 scanner. To quantify the anisotropy, the intensity was integrated at the arc-shaped (0 0 2) peak within 20°<2θ<30° [13, 24]. Herman's orientation factor (HOF) was calculated to describe the alignment fraction following [26]

$$f(HOF) = 1 - \frac{3}{2} <\sin^2\varphi> \qquad (2)$$

$$<\sin^2\varphi> = \frac{\int_{0°}^{90°} I(\varphi)\sin^2\varphi\cos\varphi d\varphi}{\int_{0°}^{90°} I(\varphi)\cos\varphi d\varphi} \qquad (3)$$

## 3. Persistent Homology

This section introduces an alignment index $\zeta$ derived from TDA, which is a scalar that quantifies the bundle alignment fraction in a complex array of orientations, as typically observed in SEM images of CNT materials (Figure 2a). As an example, we chose a local circular window (circled in red) and applied the Canny edge detector [27] to generate a binary image (Figure 2b), denoted as $X$ in the following discussion. Comparing the pre-processed binary image and the original SEM image, we see that the CNT bundle orientation information is well-maintained in the pre-processing stage, ensuring the accuracy of the following analysis based on the binary image $X$. Persistent homology was then applied, and the corresponding barcode expressions were used to develop a measure $V(X, \theta)$ for the CNT bundle alignment fraction in array X at the given direction $\theta$. Although $X$ may be a coarse and often noisy representation of the CNT network, the persistent homology barcodes still provide a robust and effective quantitative analysis of the bundle orientation in a prescribed direction. This is one of the strengths of the TDA method, as it does not require a high-resolution image or a fine edge segmentation.

To understand how the barcode representations were obtained from the output $X$ of the Canny edge detector, we simplified the entangled CNT network structures from the pre-processed binary image (Figure 2b) into the array comprising five CNT bundles shown in different colors in Figure 2c. The total number of components is known as the "zeroth Betti number" of $X$ and denoted as $b_0(X)$. In this case, $b_0(X) = 3$, as the red and yellow bundles were connected, as were the green and cyan ones. For any direction determined by angle $\theta$ ($0 \leq \theta \leq \pi$, measured from the positive x-axis), $V(X, \theta)$ quantified the total elongation of the three components in that direction: larger $V$ values indicated higher CNT alignment degree at angle $\theta$.

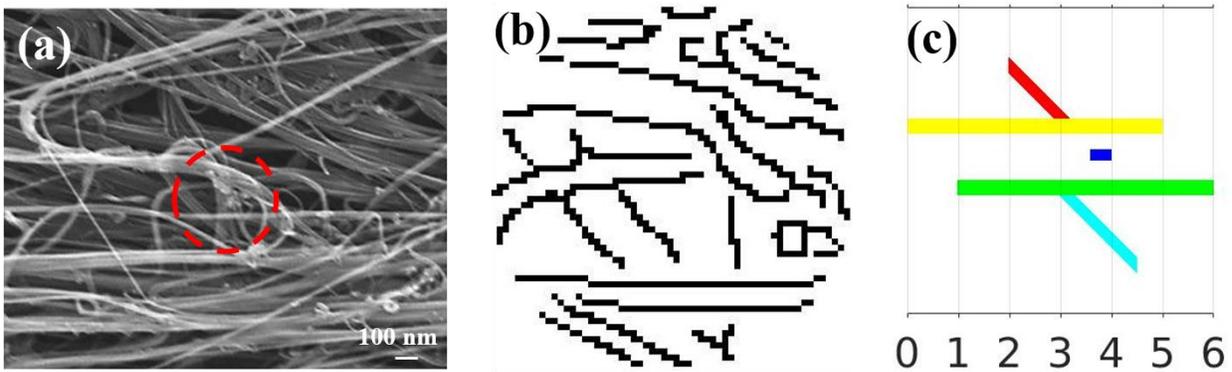

Figure 2. (a) A typical SEM image of the 30%-stretched CNT sheets (adapted from [26]); (b) the corresponding binary image as the output of the Canny edge detector; (c) an illustration of the simplified array.

### 3.1 Persistent Homology Barcodes

Without losing generality, after a rotation, we may assume that $\theta = 0$ indicating the horizontal scanning direction. Let $X_t = \{(x, y) \in X,\ x \leq t\}$ for any $0 \leq t \leq 6$, where $X_t$ represents the part of $X$ confined in the band between $x = 0$ and $x = t$. The scanning was conducted orthogonally from left to right, with $X_t$ becoming progressively larger as $t$ grew with $X_t = X$ at

$t = 6$. Following the standard TDA terminology, we referred to $\{X_t\}_{0 \leq t \leq 6}$ as the sub-level set filtration of X, shown in Figure 2(a). Similarly, when the orthogonal scanning was completed from right to left, the sequence $X^t = \{(x, y) \in X,\ x \geq t\}$ was denoted as the super-level set filtration (Figure 3(b)). The resulting continuous yet nested sequences demonstrated the appearance, merging, and bifurcation of the branches along the filtration.

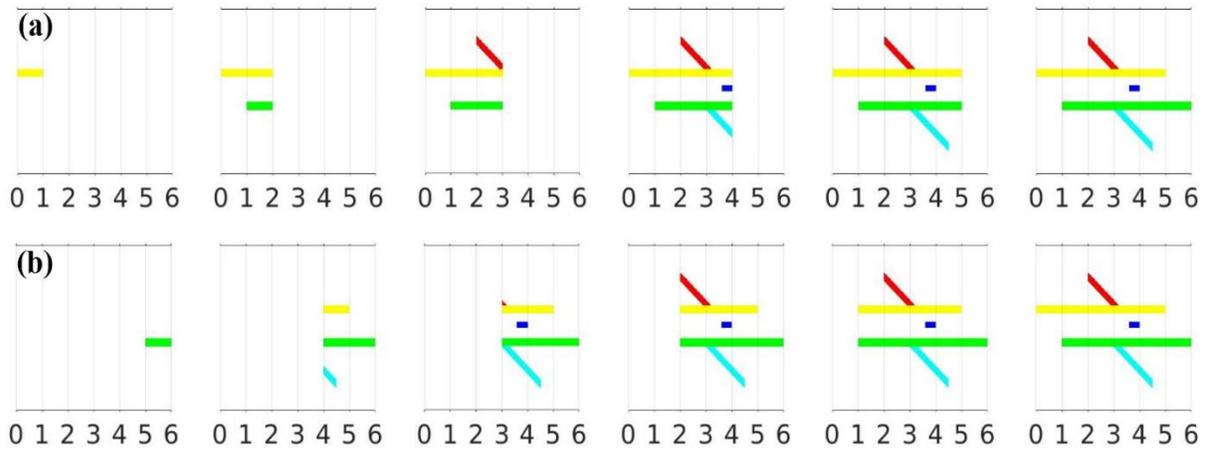

Figure 3. Various stages of the (a) sub-level and (b) super-level set filtrations of X.

Derived from the sub- and super-level filtrations, Figure 4 presents the persistence barcodes for $\{X_t\}_{0 \leq t \leq 6}$ and $\{X^t\}_{0 \leq t \leq 6}$, respectively. The barcodes can be visualized as collections of bars or intervals showing the evolution of the filtration from a horizontal perspective. The number of bars over $t$ is equivalent to the number of components in $X_t$ accordingly: for $0 \leq t < 1$, only one component was observed, resulting from the yellow bundle shown in Figure 2(c); for $t = 1$, a new green bar was born due to the emergence of the green branch, followed by the birth of a red bar at $t = 2$; for $t = 3$, the red bundle merged with the yellow one, which decreased the number of components to two. A more rigorous mathematical explanation for this convention may be found in [28, 29]. Note that the cyan branch of $X$ never contributed a new bar since it was always connected to the green one for any $t \geq 3$. Similarly, the barcode for the super-level set filtration of $X$ was constructed as illustrated in Figure 4(b). It is worth of mentioning that the color associated with the corresponding branches in $X$ is only for visualization purposes. Barcodes, in general, do not provide that information.

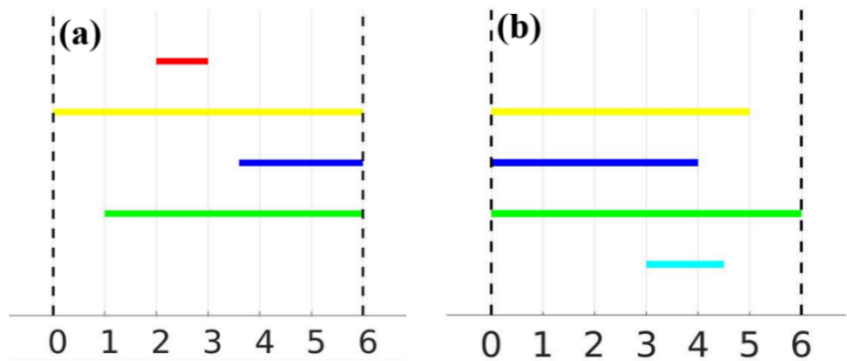

Figure 4. Barcodes for the sub-level and super-level set filtrations of array $X$: (a) $\{X_t\}$; (b) $\{X^t\}$.

**3.2 Orientation and Alignment**

To what extent do the barcodes capture the total spread $l$ of the branches in the direction $\theta$? The barcodes for $\{X_t\}$ and $\{X^t\}$ are denoted as B1 and B2, respectively. Note that the lengths of the red bar in B1 and the cyan bar in B2 precisely record the horizontal spread of the red and cyan branches of $X$. However, the other branches are not as clear such that we can directly recover the horizontal spread. For the yellow branch as an example, assuming $l_1$ and $l_2$ are the lengths in B1 and B2, respectively, and $l_0$ is the horizontal spread of the yellow branch, it is simple to verify that

$$l_1 + l_2 = l_0 + 6 \qquad (4)$$

where 6 is the width of the smallest vertical band that contains $X$. The same applies to the blue and green branches. However, in real-life applications, pairing the intervals from the sub- and super-level filtrations would be almost impossible as the barcodes can be complex yet not colored. In order to determine the total spread $V(X,\theta)$ of the branches in $X$ at the direction $\theta$, the summation of all the bar lengths in B1 and B2 were determined and the excess over the total spread $V(X,\theta)$ was $6b_0(X)$. Hence, if B1 was comprised of $m$ intervals $I_i$, $1 \leq i \leq m$ and B2 of n intervals $J_j$, $1 \leq j \leq n$, then

$$V(X,\theta) = \sum_{i=1}^{m} l(I_i) + \sum_{j=1}^{n} l(J_j) - 6b_0(X) \qquad (5)$$

Let the tightest band containing $X$ and orthogonal to the direction $\theta$ (for convenience, rotated onto the x-axis) be bound by the lines $x = m$ and $x = M$, $m < M$. Then, we define

$$V(X,\theta) = \sum_{i=1}^{m} l(I_i) + \sum_{j=1}^{n} l(J_j) - b_0(X)(M - m) \qquad (6)$$

where $I_i$ and $J_j$ are the intervals comprising the barcodes for the sub-level and super-level set filtrations of $X$ along the direction $\theta$, respectively. Foundational work on persistent homology and its barcode expression ensures that this index is robust to small disturbances of $X$ [22, 30].

Under circumstances where the CNT bundles are perfectly aligned along the direction $\theta$, the total spread $V(X,\theta)$ is maximized where that of the orthogonal direction $\theta^\perp$ is negligible, $|\theta^\perp - \theta| = \pi/2$. The alignment index $\zeta$ is then be defined as

$$\zeta = \frac{V(X,\theta_{max}) - V(X,\theta_{max}^\perp)}{V(X,\theta_{max})} \qquad (7)$$

Clearly, $0 \leq \zeta \leq 1$, with $\zeta = 1$ indicating perfect alignment.

## 4. RESULTS AND DISCUSSIONS

### 4.1 The CNT alignment validation

While the bundle alignment could be determined qualitatively with the increased strain from 0% to 40% in the SEM images shown in Figure 5 (a)-(e), a quantitative image analysis algorithm is vital for demonstrating the effectiveness and accuracy of the alignment index $\zeta$ in capturing the orientation. Applying the Canny edge detector to the circular windows of SEM images, the corresponding blue curves, shown on the second row (Figure 5 (f)-(j)), depict the value of $V(X, \theta)$ as a function of $\theta$ in a polar coordinate system, while the red segments highlight the directions that maximize the total spread. The curves were plotted in the full range $\theta \in [0, 2\pi]$. Since $\theta$ and $\theta + \pi$ represent the same directions, the plots are symmetric with respect to the origin.

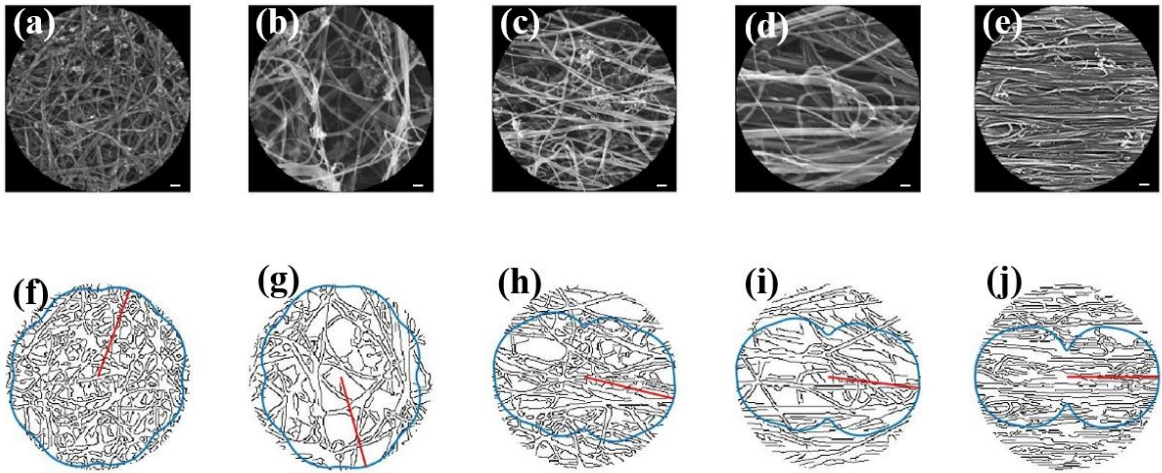

Figure 5. (a)-(e) The original SEM images for the 0%, 10%, 20%, 30% and 40% stretched CNTs, respectively, under the accelerating voltage of 10 kV at the magnification of ×50,000. Scale bar: 100 nm. Adapted from [26]; (f)-(j) the output of the Canny edge detector, with the red segments showing the preferred direction and the blue curves for the evolution of $V(X, \theta)$;

For validation purposes, the TDA results were compared to the conventional characterization methods, including polarized Raman spectroscopy and X-Ray Scattering, shown in Figure 6(a). Since the degree of alignment increased with respect to the applied strain and lies between 0 and 1, we fitted a logistic function to the data points:

$$\sigma(x; a, b) = \frac{1}{1+\exp(a+bx)} \qquad (8)$$

While the polarized Raman and TDA results were fitted in Figure 6(b) and (c), the correlation coefficient $R^2$ were calculated to be 0.972 and 0.983, respectively, indicating a reliable fitting model that well predicts the CNT bundle alignment attained from both TDA and traditional polarized Raman spectroscopy.

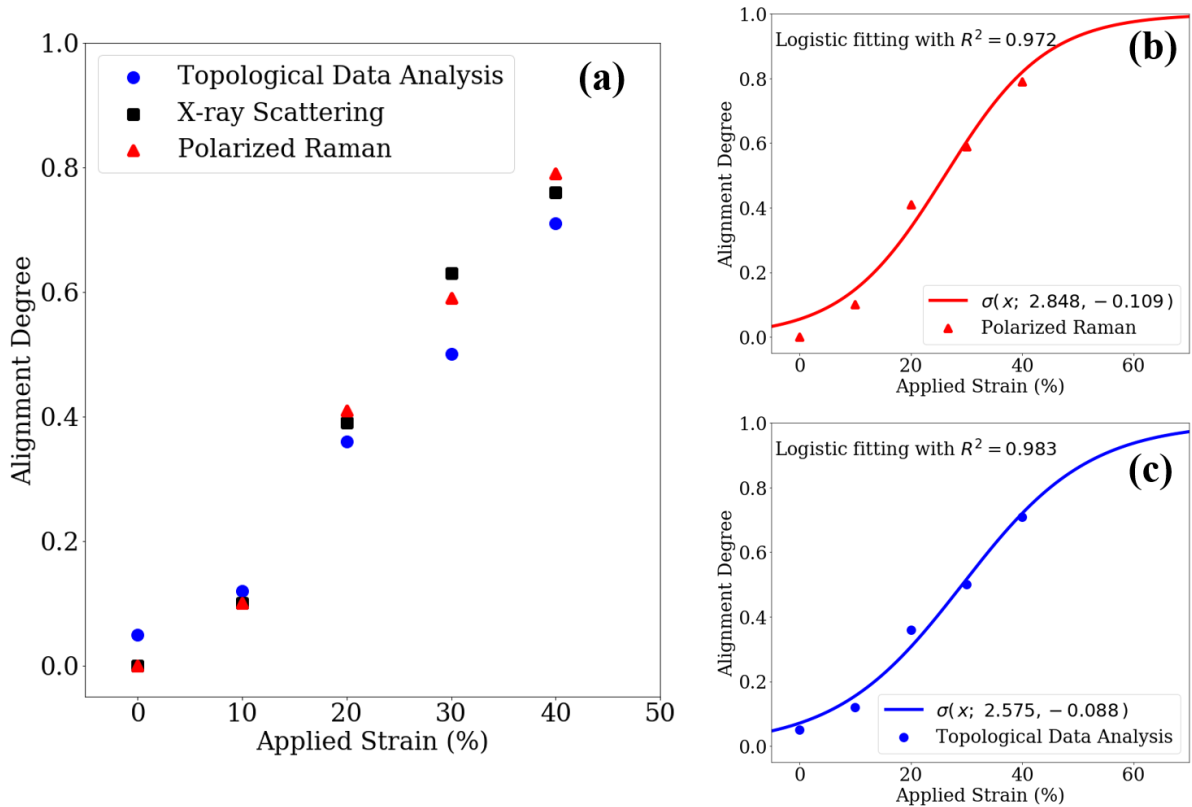

Figure 6. (a) The degree of alignment extracted from the polarized Raman spectroscopy, the X-ray scattering, and the topological data analysis; the logistic fitting curves of (b) polarized Raman Spectroscopy and (c) TDA.

Furthermore, Figure 6 shows a good match comparing the TDA results and conventional characterization results (i.e., polarized Raman spectroscopy and X-ray Scattering). To quantify the correlations between all the calculated degrees of alignment, the root mean squared deviations (RMSDs) was introduced to measure how far away the logistic models for polarized Raman spectroscopy and TDA method differed from each other over the applied strain ranging 0% ~ 100%:

$$RMSD = \sqrt{\frac{1}{100} \int_0^{100} (\sigma_1(x) - \sigma_2(x))^2 \, dx} \qquad (9)$$

which equals 0.05, indicating that these two models were very close to each other, hence the reliability of the proposed TDA method.

The minor differences could be explained by the various penetration depths or mathematical formulations. Among all three experimental characterization methods, SEM [31] and polarized Raman spectroscopy [32] were widely accepted as surface-sensitive techniques. Depending on the accelerating voltages and sample nature (i.e. CNT diameter and volume fraction), the penetration depth of SEM is estimated to be less than 100 nm due to the shallow escape depths of the secondary electrons [16], while the polarized Raman spectroscopy has been reported to penetrate deeper [33, 34]. As the alignment distribution along the through-thickness direction may be inhomogeneous, the varying penetration depths could play as a significant impacting factor in the quantitative alignment analysis. On the other hand, X-ray scattering has been deemed as a penetrative technique, but the carbonaceous impurities tend to interfere or modify the X-ray scattering patterns [35, 36], which could also lead to the variations in the calculated alignment index.

### 4.2 Effects of accelerating voltages and magnifications

With the SEM technique, the acceleration voltage empowers the electrons to penetrate the sample. Therefore, higher accelerating voltages would likely result in a more penetrative electron beam, which would reveal more internal CNT bundles and resultant intersections for topological data analysis. Therefore, the Canny edge detection would return more chopped edges along each bundle. As the expansion of a single long edge would be quite close to the expansion of several small interrupted pieces, our proposed strategy should still provide trustworthy results. Similarly, the magnifications where the SEM images were captured determine the level of detail, as the characteristic features observed could vary at different length-scales (i.e. individual CNTs and CNT bundles) [16] and the number of visible CNT bundles may also be impacted. We have investigated these factors by acquiring images of the 20% stretched samples using different accelerating voltages and magnification scales at a fixed location before applying the TDA methodology. In Figure 7, consistent degrees of alignment and directions when applying various imaging parameters were observed, and the negligible variations (<1%) once again demonstrated the reliability and robustness of the proposed TDA algorithms.

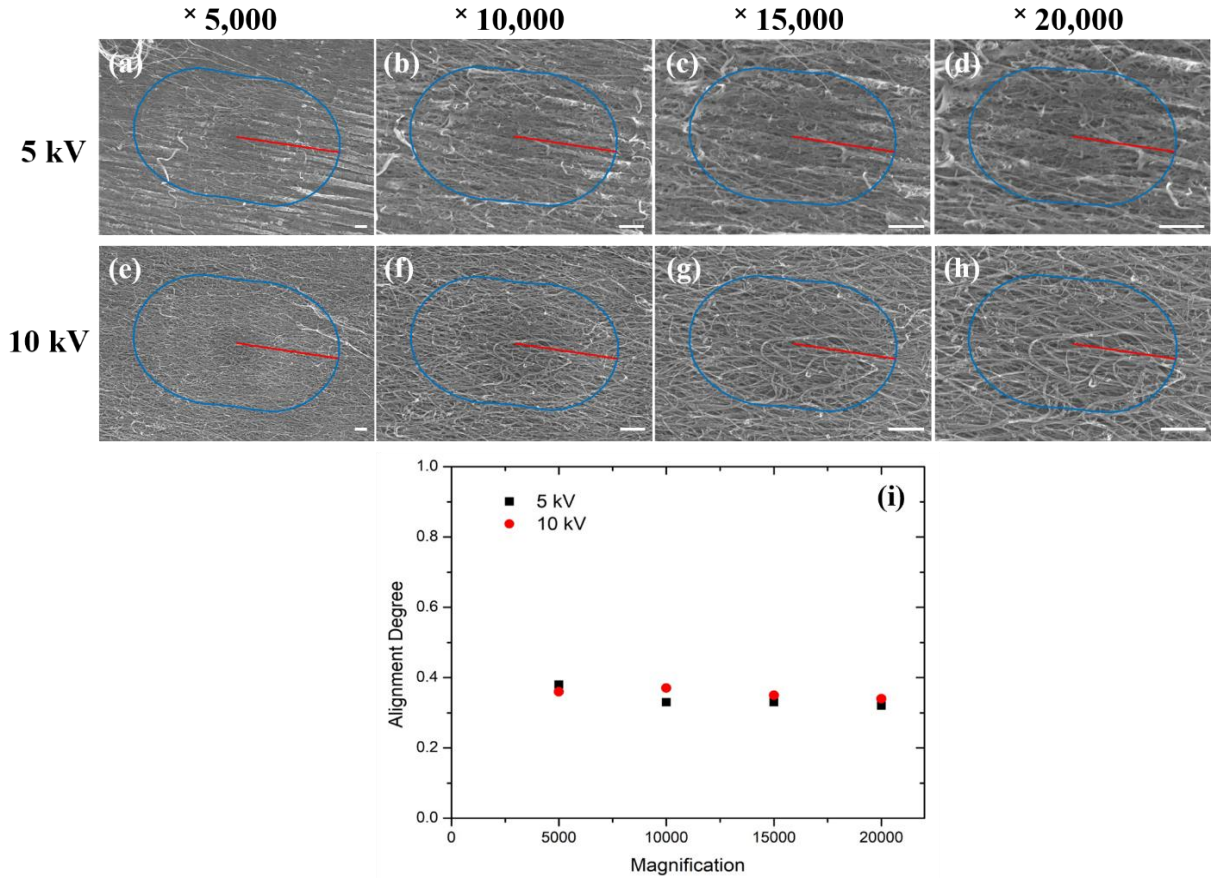

Figure 7. (a)-(d) SEM images of the 20% stretched CNTs under the acceleration voltage of 5 kV with the magnifications of ×5,000, ×10,000, ×15,000, and ×20,000, respectively; (e)-(h) SEM images of the 20% stretched CNTs under the acceleration voltage of 10 kV with the magnifications of ×5,000, ×10,000, ×15,000, and ×20,000, respectively, respectively; (i) the calculated alignment degree. Scale bar:1 μm.

## 5. CONCLUSIONS

We have developed an innovative yet simple approach to efficiently detect and map the carbon nanotube bundle alignment using the topological data analysis (TDA) based on the SEM images of thin CNT sheet materials. The CNT bundle extensions in certain directions were summarized in an algebraic way and expressed as visible barcodes, which were then calculated and converted into the total spread function $V(X, \theta)$, from which the argument value $\theta_{max}$ indicates the preferred alignment direction at the maximum spread. The alignment index, $\zeta$, was subsequently defined by tracking the spread changes in orthogonal directions, $0 \leq \zeta \leq 1$, with $\zeta = 1$ indicating perfect alignment.

To validate the proposed methodology, quantitative comparisons were made with the Herman's orientation factors (HOFs) obtained from the polarized Raman spectroscopy and the wide-angle X-ray scattering on the mechanically stretched CNT sheets with the stretching ratio ranging from 0-40%. With the good alignment fraction agreement our TDA methodology demonstrated when compared with the conventional characterizations, the deviations could be explained by the different penetration depths intrinsic of the characterizations. Additionally, the proposed approach exhibited good flexibility and robustness as the choice of SEM parameters including acceleration voltages and magnifications, which did not immediately impact the calculated alignment index $\zeta$.

While the methodology has only been tested on the aligned CNT sheets, this work may provide an alternative perspective to detect the alignment of polymer nanofibers where the Raman spectroscopy cannot provide sufficient information. We also believe that this fast detection technique could potentially be applicable in the CNT/polymer composite systems.

## ACKNOWLEDGEMENT

The authors would like to extend genuine appreciation to Mr. Frank Allen for his involvement in the language editing and proofreading. This work was partially supported by United States Air Force Office of Scientific Research (AFOSR) contract No. FA9550-17-1-0005 and National Science Foundation grant DMS-1722995.

# REFERENCE

[1] Marcelo Motta, Ya-Li Li, Ian Kinloch, A. Windle, Mechanical Properties of Continuously Spun Fibers of Carbon Nanotubes, Nano Letters 5(8) (2005) 1529-1533.
[2] Q.W. Li, Y. Li, X.F. Zhang, S.B. Chikkannanavar, Y.H. Zhao, A.M. Dangelewicz, L.X. Zheng, S.K. Doorn, Q.X. Jia, D.E. Peterson, P.N. Arendt, Y.T. Zhu, Structure-Dependent Electrical Properties of Carbon Nanotube Fibers, Advanced Materials 19(20) (2007) 3358-3363.
[3] L. Dong, J.G. Park, B.E. Leonhardt, S. Zhang, R. Liang, Continuous Synthesis of Double-Walled Carbon Nanotubes with Water-Assisted Floating Catalyst Chemical Vapor Deposition, Nanomaterials (Basel) 10(2) (2020).
[4] Y. Han, X. Zhang, X. Yu, J. Zhao, S. Li, F. Liu, P. Gao, Y. Zhang, T. Zhao, Q. Li, Bio-Inspired Aggregation Control of Carbon Nanotubes for Ultra-Strong Composites, Sci Rep 5 (2015) 11533.
[5] K. Iakoubovskii, Techniques of aligning carbon nanotubes, Central European Journal of Physics 7(4) (2009).
[6] J.E. Fischer, W. Zhou, J. Vavro, M.C. Llaguno, C. Guthy, R. Haggenmueller, M.J. Casavant, D.E. Walters, R.E. Smalley, Magnetically aligned single wall carbon nanotube films: Preferred orientation and anisotropic transport properties, Journal of Applied Physics 93(4) (2003) 2157-2163.
[7] D. Wang, P. Song, C. Liu, W. Wu, S. Fan, Highly oriented carbon nanotube papers made of aligned carbon nanotubes, Nanotechnology 19(7) (2008) 075609.
[8] T.Q. Tran, Z. Fan, P. Liu, S.M. Myint, H.M. Duong, Super-strong and highly conductive carbon nanotube ribbons from post-treatment methods, Carbon 99 (2016) 407-415.
[9] Q. Cheng, J. Bao, J. Park, Z. Liang, C. Zhang, B. Wang, High Mechanical Performance Composite Conductor: Multi-Walled Carbon Nanotube Sheet/Bismaleimide Nanocomposites, Advanced Functional Materials 19(20) (2009) 3219-3225.
[10] Q. Cheng, B. Wang, C. Zhang, Z. Liang, Functionalized carbon-nanotube sheet/bismaleimide nanocomposites: mechanical and electrical performance beyond carbon-fiber composites, Small 6(6) (2010) 763-7.
[11] Y. Han, S. Li, F. Chen, T. Zhao, Multi-scale alignment construction for strong and conductive carbon nanotube/carbon composites, Materials Today Communications 6 (2016) 56-68.
[12] B. Han, X. Xue, Y. Xu, Z. Zhao, E. Guo, C. Liu, L. Luo, H. Hou, Preparation of carbon nanotube film with high alignment and elevated density, Carbon 122 (2017) 496-503.
[13] S. Li, J.G. Park, Z. Liang, T. Siegrist, T. Liu, M. Zhang, Q. Cheng, B. Wang, C. Zhang, In situ characterization of structural changes and the fraction of aligned carbon nanotube networks produced by stretching, Carbon 50(10) (2012) 3859-3867.
[14] Benjamin N. Wang, Ryan D. Bennett, Eric Verploegen, Anastasios J. Hart, R.E. Cohen, Quantitative Characterization of the Morphology of Multiwall Carbon Nanotube Films by Small-Angle X-ray Scattering, Journal of Physical Chemistry B 111 (2007) 5859-5865.
[15] W. Li, H. Zhang, C. Wang, Y. Zhang, L. Xu, K. Zhu, S. Xie, Raman characterization of aligned carbon nanotubes produced by thermal decomposition of hydrocarbon vapor, Applied Physics Letters 70(20) (1997) 2684-2686.
[16] E. Brandley, E.S. Greenhalgh, M.S.P. Shaffer, Q. Li, Mapping carbon nanotube orientation by fast fourier transform of scanning electron micrographs, Carbon 137 (2018) 78-87.

[17] G. Carlsson, Topology and Data, American Mathematical Society 46(2) (2009) 255-308.
[18] L. Wasserman, Topological Data Analysis, Annual Review of Statistics and Its Application 5(1) (2018) 501-532.
[19] D. Cohen-Steiner, H. Edelsbrunner, J. Harer, Stability of Persistence Diagrams, Discrete & Computational Geometry 37(1) (2006) 103-120.
[20] J.L. Nielson, J. Paquette, A.W. Liu, C.F. Guandique, C.A. Tovar, T. Inoue, K.A. Irvine, J.C. Gensel, J. Kloke, T.C. Petrossian, P.Y. Lum, G.E. Carlsson, G.T. Manley, W. Young, M.S. Beattie, J.C. Bresnahan, A.R. Ferguson, Topological data analysis for discovery in preclinical spinal cord injury and traumatic brain injury, Nat Commun 6 (2015) 8581.
[21] D. Taylor, F. Klimm, H.A. Harrington, M. Kramar, K. Mischaikow, M.A. Porter, P.J. Mucha, Topological data analysis of contagion maps for examining spreading processes on networks, Nat Commun 6 (2015) 7723.
[22] M. Saadatfar, H. Takeuchi, V. Robins, N. Francois, Y. Hiraoka, Pore configuration landscape of granular crystallization, Nat Commun 8 (2017) 15082.
[23] M. Li, M.H. Frank, V. Coneva, W. Mio, D.H. Chitwood, C.N. Topp, The Persistent Homology Mathematical Framework Provides Enhanced Genotype-to-Phenotype Associations for Plant Morphology, Plant Physiol 177(4) (2018) 1382-1395.
[24] A. Zomorodian, G. Carlsson, Computing Persistent Homology, Discrete & Computational Geometry 33(2) (2004) 249-274.
[25] R.D. Downes, A. Hao, J.G. Park, Y.-F. Su, R. Liang, B.D. Jensen, E.J. Siochi, K.E. Wise, Geometrically constrained self-assembly and crystal packing of flattened and aligned carbon nanotubes, Carbon 93 (2015) 953-966.
[26] R. Downes, S. Wang, D. Haldane, A. Moench, R. Liang, Strain-Induced Alignment Mechanisms of Carbon Nanotube Networks, Advanced Engineering Materials 17(3) (2015) 349-358.
[27] J. J. Hermans, P. H. Hermans, D. Vermaas, A. Weidinger, Quantitative Evaluation of Orientation in Cellulose Fibres from the X-Ray Fibre Diagram, Recueil des Travaux Chimiques des Pays-Bas 65 (1945).
[28] J. Canny, A Computational Approach to Edge Detection, IEEE Transactions on Pattern Analysis and Machine Intelligence 8(6) (1986) 679-698.
[29] Gunnar Carlsson, Afra Zomorodian, Anne Collins, L. Guibas, Persistence Barcodes for Shapes, International Journal of Shape Modeling 11 (2005) 149-188.
[30] R. Ghrist, Barcodes: the Persistent Topology of Data, BULLETIN (New Series) OF THE AMERICAN MATHEMATICAL SOCIETY 45(1) (2008) 61-75.
[31] Joseph I. Goldstein, Dale E. Newbury, Joseph R. Michael, Nicholas W.M. Ritchie, John Henry J. Scott, D.C. Joy, Scanning Electron Microscopy and X-ray Microanalysis, 4 ed., Springer, New York, NY, 2018.
[32] J. Song, C. Yang, H. Hu, X. Dai, C. Wang, H. Zhang, Penetration depth at various Raman excitation wavelengths and stress model for Raman spectrum in biaxially-strained Si, Science China Physics, Mechanics and Astronomy 56(11) (2013) 2065-2070.
[33] M.S. Dresselhaus, G. Dresselhaus, R. Saito, A. Jorio, Raman spectroscopy of carbon nanotubes, Physics Reports 409(2) (2005) 47-99.
[34] M.S. Dresselhaus, A. Jorio, A.G. Souza Filho, R. Saito, Defect characterization in graphene and carbon nanotubes using Raman spectroscopy, Philos Trans A Math Phys Eng Sci 368(1932) (2010) 5355-77.

[35] Z.Q. Li, C.J. Lu, Z.P. Xia, Y. Zhou, Z. Luo, X-ray diffraction patterns of graphite and turbostratic carbon, Carbon 45(8) (2007) 1686-1695.
[36] H. Fujimoto, Theoretical X-ray scattering intensity of carbons with turbostratic stacking and AB stacking structures, Carbon 41(8) (2003) 1585-1592.